\documentclass[conference]{IEEEtran}
\ifCLASSINFOpdf
  \usepackage[pdftex]{graphicx}
  \DeclareGraphicsExtensions{.pdf,.jpeg,.png}
\else
  \usepackage[dvips]{graphicx}
\fi

\usepackage{pgf}

\definecolor{darkgreen}{rgb}{0,0.7,0}

\newif\ifdraft
\drafttrue
\ifdraft
 \newcommand{\katznote}[1]{ {\textcolor{blue} { ***Dan:   #1 }}}
 
 \newcommand{\kriedernote}[1]{ {\textcolor{darkgreen}  { ***Scott:   #1 }}}
 \newcommand{\note}[1]{ {\textcolor{red}    {\bf #1 }}}
\else
 \newcommand{\katznote}[1]{}
 \newcommand{\kriedernote}[1]{}
 \newcommand{\note}[1]{}
\fi

\begin{document}
%
\title{Reusability in Science: From Initial User Engagement
               to Dissemination of Results}


\author{\IEEEauthorblockN{Ketan Maheshwari\IEEEauthorrefmark{1},
David Kelly\IEEEauthorrefmark{1},
Scott J. Krieder\IEEEauthorrefmark{2},
Justin M. Wozniak\IEEEauthorrefmark{1},\\
Daniel S. Katz\IEEEauthorrefmark{3},
Mei Zhi-Gang\IEEEauthorrefmark{4},
Mainak Mookherjee\IEEEauthorrefmark{5}
\IEEEauthorblockA{
\IEEEauthorrefmark{1}MCS Division, Argonne National Laboratory}
\IEEEauthorrefmark{2}Department of Computer Science, Illinois Institute of Technology\\
\IEEEauthorrefmark{3}Computation Institute,  University of Chicago \& Argonne National Laboratory\\
\IEEEauthorrefmark{4}Nuclear Engineering Division, Argonne National Laboratory\\
\IEEEauthorrefmark{5}Department of Earth and Atmospheric Sciences, Cornell University}
}

\maketitle

\begin{abstract}
Effective use of parallel and distributed computing in science depends upon multiple interdependent entities and activities that form an ecosystem.
Active engagement between application users and technology catalysts is a crucial activity that forms an integral part of this ecosystem.
Technology catalysts play a crucial role benefiting communities beyond a single user group.
An effective user-engagement, use and reuse of tools and techniques has a broad impact on software sustainability. 
From our experience, we sketch a life-cycle for user-engagement activity in scientific computational environment and posit that application level reusability promotes software sustainability. 
We describe our experience in engaging two user groups from different scientific domains reusing a common software and configuration on different computational infrastructures.  
\end{abstract}


\IEEEpeerreviewmaketitle


\section{Introduction}
Domain scientists often have limited time to investigate the capabilities that
a large scale computing and data-handling infrastructure combined with a high
performance software framework could bring to their scientific activities.
Technology catalysts help speed up the tedious process of organizing scientific
computations such that they can be easily mapped onto computational
infrastructure. However, this is an iterative process and not free of pitfalls.
The source of these pitfalls can be the scientific process itself or a mismatch
in technical requirements mapping to computational infrastructures. 

This presents a challenge: enabling effective reuse of existing user-engagement
patterns and related products for a new scientific user. If this challenge can
be met, it could lead to a considerable acceleration in the process of
conceptualizing, describing, defining, deploying, and executing scientific
experiments. 
Reuse of data and software libraries is fairly common. Reuse of enabled
applications across scientific domains is not as common. A successful execution
of such applications might require tuning specific to the application
requirements. However, with familiarization from previous engagements, much of
this process can be expedited. 

%

A widely reused system has a higher sustainability as a community supports its
maintenance. Enabling application level reuse promotes sustainability of the entire
ecosystem of modern science.
In this experience paper, we report on the following:

\begin{enumerate}
  \item Experience in scientific community engagement describing activities
      performed at different levels in order to support scientific users with
      applications deployed onto new, larger and faster systems.
  
  \item A sketch and demonstration the elements of a successful scientific
      application deployment cycle. 
  
  \item Enhancements of an enabling software framework based on user feedback
      resulting in a software with improved usability.
\end{enumerate}

The remainder of this paper is organized as follows.  In
Section~\ref{section:cycle}, we describe the user engagement cycle that
provides a context to the human aspects of our work.  In
Section~\ref{section:users}, we describe the applications on which our
experience is based and in which we apply software reuse techniques.  In
Section~\ref{section:infrastructure}, we describe the hardware and software
complexities that make software maintenance strategies important.  In
Section~\ref{section:related}, we summarize some related work, and in
Section~\ref{section:conclusion} we offer concluding remarks. 

\section{User Engagement Cycle}
\label{section:cycle}

\begin{figure}
  \begin{center}
      \includegraphics[width=7cm]{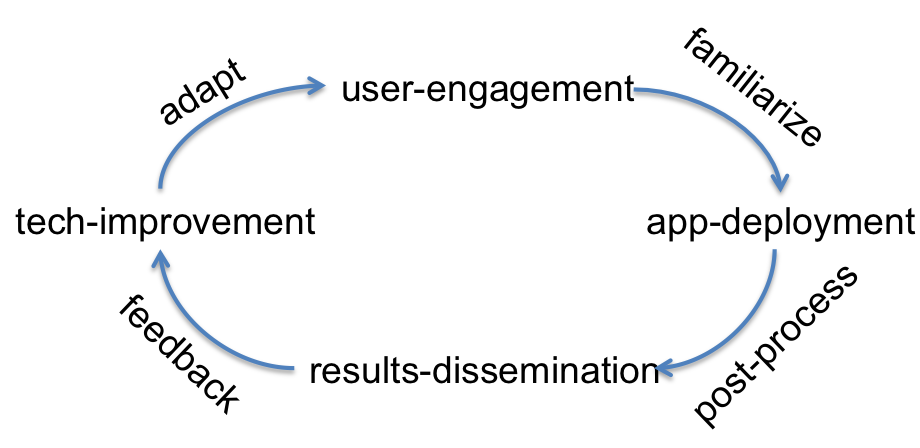}

  \caption{Activities and transitions in user engagement cycle.
    \label{fig:user}
  }
\end{center}
\end{figure}

User engagement with technical catalysts is a complex social process
that differs with respect to institutions, culture, and technical
practices. A distillation of key points in the process is diagrammed
in Figure~\figurename~\ref{fig:user}. The cycle consists of four
phases and transition activities between these phases. It starts with
user-catalyst communication involving familiarizing the domain science
and technology from the user side and computational tools and
technology from the catalyst side with each other. The result is a
map of scientific tools onto the computational infrastructure.  The
next phase is application deployment and execution on these resources.
Enabling software tools are put into practice in this phase. This ends
in post-processing activity involving collection, pruning, structuring
of results.  The next phase is the results-dissemination phase. During
this phase, the science results obtained in the previous activity are
presented to the scientific user. Adaptive users often do further
analysis themselves and trigger a next stage via their feedback. The
feedback results in technological improvements and adjustments to
better suite the requirements of application on hand.

In our experience, demonstrating the results of engagement with one
user-group to the other 
has significantly inspired, and aroused interests in adapting modern high
performance computational techniques. This method led to a productive
cycle of defining and executing experiments which has benefited both
groups. This affirms the role of technology catalysts by enabling the
use of cross-application knowledge. 

From a software sustainability perspective, identifying software
patterns and applying them into new applications is an opportunity to
maximize the initial investment in the software research, as well as
speed progress in later projects. An example of this process, based
on application experience, is described in the next section. 

\section{User Groups}
\label{section:users}

In this section we describe the two user groups we engaged with in
this work.  First, the Mineral Physics Group at the Department of
Earth and Atmospheric Science at Cornell University, and second, the
Material Science Group at Argonne National Laboratory. As described
below, the application areas differ greatly (geological research
vs. nuclear energy), yet the underlying software tools and challenges
have much to gain from software reuse.

\subsection{Mineral Physics}
Rocks and minerals exposed at the surface of the Earth are in constant
interaction with the overlying hydrosphere. Over geological time scales, this
leads to stabilization of hydrous mineral phases such as
serpentine~\cite{serpentine}. These lock up oceanic water. These minerals are
then dragged down along subduction zones. This process has a long term
influence on the global sea levels. In order to understand how much water is
dragged into the deep interior of the Earth, one must have a better
understanding of the energetics and thermo-elasticity of these hydrous phases
stable at subduction zone conditions.

We use ab-initio simulations based on density functional theory to understand
the energetics and thermo-elasticity of hydrous mineral phases relevant to the
subduction zone conditions. The mineral phases stable at the subduction zone
conditions often have lower space-group symmetry and a large number of atoms in
their unit cell. The compute requirements of simulations are directly
proportional to the number of atoms in a unit cell of a mineral. Consequently
the simulations are computationally intensive. 



\begin{figure}
    \begin{center}
    \includegraphics[width=7cm]{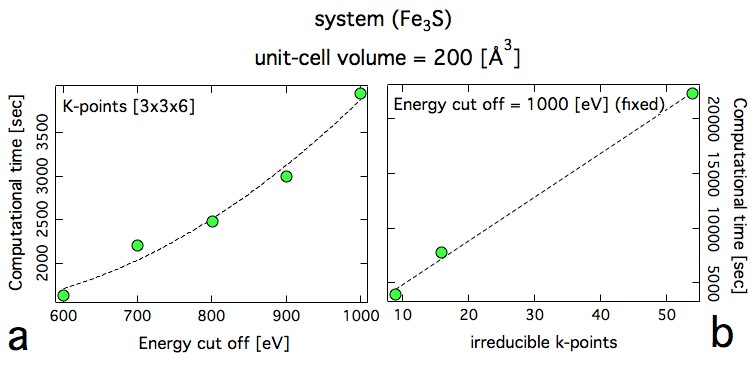}
    \caption{Constant volume Fe${_3}$S unit-cell experiments: (a) Plot of computational time as a function of cut-off energy for a constant k-point mesh of 3x3x6; (b) plot of computational time vs. irreducible k-points for a constant cut-off energy.\label{fig:timeline}}
\end{center}
\end{figure}

For our research we use the Vienna Ab-initio Simulation Package
(VASP)~\cite{vasp} software suite. VASP has been quite successful in predicting
band structure, ground state energy, and physical properties including
elasticity. From our experience using institutional resources, we find that a 
simulation for a system-size of approximately 100 atoms requires 24
hours of CPU time. For a single mineral phase, we typically need 40-50
simulations to determine the full elastic constant tensor and its pressure
dependence. One comprehensive analysis on a mineral phase requires a
combination of parameters to test for convergence of ground state energy. This
requires intensive use of computational resources. In addition, we explore
solid-solution thermodynamics across various end-member chemistry of the
mineral phases.

The initial, relatively small study with 32 atoms was conducted in two stages. First, a
determination of volume at which the energy is minimum was conducted. 
From this, the unit cell volume of 200 angstroms was chosen for further
analysis over a constant 3-D volumetric mesh (\figurename~\ref{fig:timeline}-a)
and constant energy (\figurename~\ref{fig:timeline}-b).
\figurename~\ref{fig:timeline}-b shows that over five hours of normalized CPU
time is required for a single cell volume analysis.

\subsection{Material Science}
Cerium dioxide (CeO$_{2}$) is an important material with a wide range of
technological applications. It is used as an electrolyte in solid oxide fuel
cells (SOFCs), as a catalytic converter in the automotive industry, and as a
model material for PuO$_{2}$ in nuclear energy applications. It has a fluorite
structure and can develop a complex pattern of defects, depending on
temperature and oxygen pressure~\cite{matsci}. 
In order to fully understand and to be able to accurately predict the
micro-structural evolution under irradiation, elucidating the underlying
formation mechanisms of the extended defects is important.


The purpose of this study is to provide a unified view of the effect of
non-stoichiometry and temperature on the formation and evolution of nano-scaled
defects, i.e., defect clusters, voids and fission gas bubbles, in irradiated
Ceria using existing experimental data, multiscale modeling, and computer
simulations.

The migration barriers calculations using the DFT-NEB (Density Functional
Theory-Nudged Elastic Band) method and the defect structure evolution
calculations using molecular dynamics are computationally expensive. The
computations require a two-stage interdependent execution of VASP over the
material structure definition. The first stage performs relaxation and static
calculations to predict atomic positions. The second stage DFT-NEB calculations
rely on the structures produced by the previous stage. A few hundred thousand
CPU hours are expected for the calculations. Currently, we are using high
performance computing clusters with more than 2,000 cores connected by fast
InfiniBand.


\subsection{Computational Tools and Techniques}

While the two applications belong to different scientific domains,
their computational profile is similar. Both use the VASP software tool with
similar input data structures and computational stages. Both applications 
first run a calibration stage to find a close structural range before moving to
a final compute intensive stage. Consequently, the general computational
structure of the applications is similar, and they can be run using the
same computational tools, specifically 
the Swift~\cite{Swift_2011} parallel scripting tool. 
Because a full technical description
of Swift is out of scope for this paper, the following points summarize Swift's
capabilities.

\begin{itemize}
    \item Swift is an open-source, Apache-licensed software framework for
        distributed computing designed for scientific users. A C-like scripted
        language expresses applications as workflows.

    \item Swift makes it easy to run ordinary application programs on parallel and remote
        computers (from laptops to supercomputers).

    \item Swift works with a variety of resource managers and file transfer protocols.
         It uses ssh, PBS, Condor, SLURM, LSF, SGE, Cobalt, and Globus to
        run applications, and scp, http, ftp, and GridFTP to move data.

    \item Swift's \texttt{foreach} statement is the main workhorse of
        the language; it executes all iterations of a loop concurrently. The
        actual number of tasks executed in parallel is based on available
        resources and settable ``throttles". 
\end{itemize}

Workflows in the form of scripts result in portable and flexible
expressions of applications. For instance, running a single study
using different simulation/software packages from different vendors or
calculation approaches becomes convenient without changing the
application logic.  Additionally, portable expression of workflows
enables the use of multiple computational infrastructures, as
described in the next section. 


\section{Cyberinfrastructures and Technology}
\label{section:infrastructure}

In both of our application cases, users were introduced to new
computational infrastructures, in our case the NSF-funded XSEDE and
the Argonne Laboratory Computing Resource Center (LCRC) Blues systems.
The minor adaptation of highly portable application Swift scripts
resulted in minimal application porting delays before making use of
these systems, demonstrating software sustainability in the scripts
and other technological investments.

In the rest of this section we describe the enhancements in Swift triggered
directly or indirectly as a result of the user engagement activities.

\subsection{Configuration}
With a complex set of communication and execution methods on local and remote
systems, Swift offers a highly sophisticated, fine tuned to-the-core
configuration options to users. However, the disadvantage of this that users
often have to adjust parameters in different configuration files. To get around
these inconveniences, a unified and abstracted approach to configuration was
designed. Under this approach, a single file, structured as name-value pairs, is
employed to record diverse properties such as execution sites, data management,
application management and miscellaneous configuration options (e.g. retry
counts). Additionally, predefined templates with a spectrum of configurations
help users select one (out of the box) that works for their requirements.

\subsection{Galaxy}
While a textual representation of application and a terminal based execution
like in Swift gives users flexibility and more control, 
often a visual interface carries more appeal to users. With this requirement in
mind, we have conceptualized an integration of Swift with the Galaxy portal
environment~\cite{galaxy}. The integration offers users an interactive, visual
interface to large scale computational systems while benefiting from both
Swift's and Galaxy's strengths in the scientific community. With this
development, independent Swift applications can be turned into Galaxy
executable tools and used from within the Galaxy environment. A user dialog
enables users to enter application and high-level execution parameters such as
the target execution site on which to run the application.

\subsection{Accelerators}
Special hardware systems such as accelerators are gaining prominence in HPC
systems. The host CPU offloads work to an accelerator such as a graphics
processing unit (GPU) which relieves the CPU of precious compute cycles.
However, code reusability on these systems is essentially non-existent,
resulting in customized porting issues of codes written using specific
programming languages that are capable of targeting accelerators (e.g., CUDA,
OpenCL, OpenACC). Additionally, there are performance, portability, and
programmability issues arising from conflicting architectures (e.g., NVIDIA vs.
AMD vs. Intel Xeon Phi). One solution so far has been modular code, with
clearly defined inputs and outputs. VASP and related
software~\cite{spiga2012phigemm} have been extended to work with NVIDIA GPUs
through use of CUDA~\cite{hutchinson2012vasp}. Since computation is dependent
on architecture, codes have to be maintained for all of these architectures,
reducing portability. GeMTC (GPU Enabled Many-Task Computing) addresses these
issues from a Many-Task Computing approach, by developing a library of GPU
kernels that are callable from Swift. The effort targets the lowest level of
hardware and is working towards a maintainable and runnable
architecture-agnostic high-level software tool~\cite{kriederSC12}.

\section{Software reuse}
In this section, we show a snapshot of Swift and wrapper scripts used and reused in the two engagements.
The code below shows Swift's `app' definition for VASP call.

\begin{center}
\scriptsize

\begin{tabular}{r|l}
 1 & {\tt app\ (file\ o,\ file\ e,\ file\ outcar,\ file\ contcar)} \\
 2 & {\tt run\_vasp\ (file\ vasp\_incar,\ file\ vasp\_poscar,} \\
 3 & {\tt \ \ \ \ \ \ \ \ \ \ file\ vasp\_potcar,\ file\ vasp\_kpoints,} \\
 4 & {\tt \ \ \ \ \ \ \ \ \ \ string\ \_dir,\ string\ \_subdir)} \\
 5 & {\tt \{                                      } \\
 6 & {\tt \ \ runvasp\ @vasp\_incar\ @vasp\_poscar} \\
 7 & {\tt \ \ \ \ \ \ \ \ \ \ @vasp\_potcar\ @vasp\_kpoints} \\
 8 & {\tt \ \ \ \ \ \ \ \ \ \ \_dir\ \_subdir\ stdout=@o\ stderr=@e;} \\
 9 & {\tt \}                                      } \\

\end{tabular}
\\
\vspace{1mm}
\textbf{Swift binding to VASP program}
\end{center}

The following Swift code snippet is used for mineral physics application.
\begin{center}
\scriptsize
\begin{tabular}{r|l}
 1 & {\tt string\ dirs[]\ =\ [\ "300",\ "400",\ "500",\ ...\ ];} \\
 2 & {\tt file\ out[]<simple\_mapper;             } \\
 3 & {\tt \ \ \ \ \ \ \ \ \ \ \ location="stdouts",} \\
 4 & {\tt \ \ \ \ \ \ \ \ \ \ \ prefix="vasp",\ suffix=".out">;} \\
 5 & {\tt foreach\ dir,\ i\ in\ dirs\ \{          } \\
 6 & {\tt \ \ file\ incar<dir,"/","INCAR">;       } \\
 7 & {\tt \ \ file\ kpoints<dir,"/","KPOINTS">;   } \\
 8 & {\tt \ \ file\ potcar<dir,"/","POTCAR">;     } \\
 9 & {\tt \ \ file\ poscar<dir,"/","POSCAR">;     } \\
10 & {\tt \ \ out[i]\ =\ vasp(incar,\ kpoints,\ potcar,\ poscar);} \\
11 & {\tt \}                                      } \\
\end{tabular}
\\
\vspace{2mm}
\textbf{VASP usage by minerals physics application}
\end{center}

The following Swift code snippet is used for materials science applications respectively.
\begin{center}
\scriptsize
\begin{tabular}{r|l}
1 & {\tt string\ dirs[]\ =\ [\ "0L",\ "1L1",\ "1L3",\ ...];} \\
2 & {\tt string\ subdirs[]\ =\ ["pos0","pos1"];  } \\
3 & {\tt                                         } \\
4 & {\tt foreach\ dir\ in\ dirs\ \{              } \\
5 & {\tt \ foreach\ subdir\ in\ subdirs\ \{      } \\
6 & {\tt                                         } \\
7 & {\tt /* file declarations omitted                  } \\
8 & {\tt  for brevity  */                              } \\
9 & {\tt                                         } \\
10 & {\tt \ \ (output,\ error,\ outcar,\ contcar)\ =} \\
11 & {\tt \ \ \ \ runvasp(incar\_relax,\ poscar,\ potcar,} \\
12 & {\tt \ \ \ \ \ \ \ \ \ \ \ \ kpoints,\ dir,\ subdir);} \\
13 & {\tt \ \}\}                                  } \\
\end{tabular}
\\
\vspace{2mm}
\textbf{VASP usage by materials science application}
\end{center}

The following code shows the common wrapper script used.
\begin{center}
\scriptsize
\begin{tabular}{r|l}
 1 & {\tt \#!/bin/bash                            } \\
 2 & {\tt                                         } \\
 3 & {\tt incar=\$1                               } \\
 4 & {\tt poscar=\$2                              } \\
 5 & {\tt potcar=\$3                              } \\
 6 & {\tt kpoints=\$4                             } \\
 7 & {\tt dirname=\$5                             } \\
 8 & {\tt subdirname=\$6                          } \\
 9 & {\tt                                         } \\
10 & {\tt mpiexec\ -machinefile\ \$PBS\_NODEFILE\ \textbackslash} \\
11 & {\tt \ \ /blues/nfs/home/jlow/vasp/vasp.5.3/vasp} \\
12 & {\tt                                         } \\
13 & {\tt cp\ OUTCAR\ \ output/vasp-outcar-\$dirname.\$subdirname} \\
14 & {\tt cp\ CONTCAR\ output/vasp-contcar-\$dirname.\$subdirname} \\

\end{tabular}
\\
\vspace{2mm}
\textbf{Reusable shell script to launch parallel VASP under Swift}
\end{center}

\section{Related Work}
\label{section:related}
The challenges and value of developing robust software for scientific computing
has been documented in the past~\cite{bestpractice}. The idea of workflows as
enablers of reusable software is well embedded in scientific community. The
subject of reusable and sustainable tools has been addressed by the
Taverna~\cite{oinn2004taverna} and Cactus~\cite{cactus} developments. A library
of scientific workflow components, acting as a virtual shelf of applications,
has been conceived in the past. Pegasus \cite{deelman2005pegasus} is a
framework for mapping complex scientific workflows onto distributed systems.
Pegasus combined with HUBzero\cite{mclennan2010hubzero}, an online platform for
dissemination and collaboration, provides a mechanism for enabling the masses
to utilize scientific workflows \cite{mclennan13bringing}.
RunMyCode.org~\cite{runmycode} is cloud-based tool to expedite the process of
reproducing results. Researchers share their software with collaborators and
reviewers who can easily test and evaluate the work without concern for the
underlying hardware or environment. 

\section{Conclusion}
\label{section:conclusion}

In this paper, we presented our experience of engagement between
technology catalysts and scientific users. We describe the process of
application-level reusability via a typical case wherein the
commonalities between two distinct applications benefited the
respective user communities.  Additionally, we described how Swift is
a simple and effective tool for task parallel computing on
high-performance and/or distributed systems that offers unique
capabilities to promote software sustainability.

Two aspects of the experience in this work helped improve the Swift
framework: identifying usage patterns and science user feedback. We
expect the Swift usability enhancements will enable wider community
adoptation and improve ease of conducting science on multiple
infrastructures. We expect such a collaborative science to have short
and long term benefits by inspiring similar efforts in the broader
community across disciplines. Our experience demonstrates that
effective reusability of software in science is crucial and involves
much more than just the technical aspects.

\section{Acknowledgments}
This work was partially supported by the U.S. Department of Energy, under
Contract No. DE-AC02-06CH11357. Some work by DSK was supported by the National
Science Foundation, while working at the Foundation. Any opinion, finding, and
conclusions or recommendations expressed in this material are those of the
authors and do not necessarily reflect the views of the National Science
Foundation.
\bibliographystyle{IEEEtran}
\bibliography{ref}

\begin{thebibliography}{10}
\providecommand{\url}[1]{#1}
\csname url@samestyle\endcsname
\providecommand{\newblock}{\relax}
\providecommand{\bibinfo}[2]{#2}
\providecommand{\BIBentrySTDinterwordspacing}{\spaceskip=0pt\relax}
\providecommand{\BIBentryALTinterwordstretchfactor}{4}
\providecommand{\BIBentryALTinterwordspacing}{\spaceskip=\fontdimen2\font plus
\BIBentryALTinterwordstretchfactor\fontdimen3\font minus
  \fontdimen4\font\relax}
\providecommand{\BIBforeignlanguage}[2]{{%
\expandafter\ifx\csname l@#1\endcsname\relax
\typeout{** WARNING: IEEEtran.bst: No hyphenation pattern has been}%
\typeout{** loaded for the language `#1'. Using the pattern for}%
\typeout{** the default language instead.}%
\else
\language=\csname l@#1\endcsname
\fi
#2}}
\providecommand{\BIBdecl}{\relax}
\BIBdecl

\bibitem{serpentine}
M.~Mookherjee and L.~Stixrude, ``Structure and elasticity of serpentine at
  high-pressure,'' \emph{Earth and Planetary Science Letters}, vol. 279, no.
  1-2, pp. 11--19, 2009.

\bibitem{vasp}
G.~Kresse and J.~Furthm{\"u}ller, ``Software {VASP}, vienna (1999),''
  \emph{Phys. Rev. B}, vol.~54, no.~11, p. 169, 1996.

\bibitem{matsci}
\BIBentryALTinterwordspacing
D.~S. Aidhy, D.~Wolf, and A.~El-Azab, ``Comparison of point-defect clustering
  in irradiated {CeO}$_2$ and {UO}$_2$: A unified view from molecular dynamics
  simulations and experiments,'' \emph{Scripta Materialia}, vol.~65, no.~10,
  pp. 867 -- 870, 2011. [Online]. Available:
  \url{http://www.sciencedirect.com/science/article/pii/S1359646211004507}
\BIBentrySTDinterwordspacing

\bibitem{Swift_2011}
M.~Wilde, M.~Hategan, J.~M. Wozniak, B.~Clifford, D.~S. Katz, and I.~Foster,
  ``Swift: A language for distributed parallel scripting,'' \emph{Par. Comp.},
  vol.~37, pp. 633--652, 2011.

\bibitem{galaxy}
B.~Giardine, C.~Riemer, R.~C. Hardison, R.~Burhans, L.~Elnitski, P.~Shah,
  Y.~Zhang, D.~Blankenberg, I.~Albert, J.~Taylor \emph{et~al.}, ``Galaxy: {A}
  platform for interactive large-scale genome analysis,'' \emph{Genome
  research}, vol.~15, no.~10, pp. 1451--1455, 2005.

\bibitem{spiga2012phigemm}
F.~Spiga and I.~Girotto, ``{phiGEMM}: {A} {CPU-GPU} library for porting
  {Quantum} {ESPRESSO} on hybrid systems,'' in \emph{Parallel, Distributed and
  Network-Based Processing (PDP), 2012 20th Euromicro International Conference
  on}.\hskip 1em plus 0.5em minus 0.4em\relax IEEE, 2012, pp. 368--375.

\bibitem{hutchinson2012vasp}
M.~Hutchinson and M.~Widom, ``{VASP} on a {GPU}: Application to exact-exchange
  calculations of the stability of elemental boron,'' \emph{Computer Physics
  Communications}, vol. 183, no.~7, pp. 1422--1426, 2012.

\bibitem{kriederSC12}
S.~J. Krieder and I.~Raicu, ``Towards the support for many-task computing on
  many-core computing platforms,'' Doctoral Showcase, IEEE/ACM
  Supercomputing/SC, 2012.

\bibitem{bestpractice}
D.~A. Aruliah, C.~T. Brown, N.~P.~C. Hong, M.~Davis, R.~T. Guy, S.~H.~D.
  Haddock, K.~Huff, I.~Mitchell, M.~Plumbley, B.~Waugh, E.~P. White, G.~Wilson,
  and P.~Wilson, ``Best practices for scientific computing,'' \emph{CoRR}, vol.
  abs/1210.0530, 2012.

\bibitem{oinn2004taverna}
T.~Oinn, M.~Addis, J.~Ferris, D.~Marvin, M.~Senger, M.~Greenwood, T.~Carver,
  K.~Glover, M.~R. Pocock, A.~Wipat \emph{et~al.}, ``Taverna: {A} tool for the
  composition and enactment of bioinformatics workflows,''
  \emph{Bioinformatics}, vol.~20, no.~17, pp. 3045--3054, 2004.

\bibitem{cactus}
\BIBentryALTinterwordspacing
G.~Allen, T.~Goodale, F.~Löffler, D.~Rideout, E.~Schnetter, and E.~L. Seidel,
  ``Component specification in the {Cactus} {Framework}: {T}he {Cactus}
  {Configuration} {Language},'' \emph{CoRR}, vol. abs/1009.1341, 2010.
  [Online]. Available:
  \url{{http://dblp.uni-trier.de/db/journals/corr/corr1009.html}}
\BIBentrySTDinterwordspacing

\bibitem{deelman2005pegasus}
E.~Deelman, G.~Singh, M.-H. Su, J.~Blythe, Y.~Gil, C.~Kesselman, G.~Mehta,
  K.~Vahi, G.~B. Berriman, J.~Good \emph{et~al.}, ``Pegasus: {A} framework for
  mapping complex scientific workflows onto distributed systems,''
  \emph{Scientific Programming}, vol.~13, no.~3, pp. 219--237, 2005.

\bibitem{mclennan2010hubzero}
M.~McLennan and R.~Kennell, ``{HUBzero}: {A} platform for dissemination and
  collaboration in computational science and engineering,'' \emph{Computing in
  Science \& Engineering}, vol.~12, no.~2, pp. 48--53, 2010.

\bibitem{mclennan13bringing}
M.~McLennan, S.~Clark, E.~Deelman, M.~Rynge, K.~Vahi, F.~McKenna, D.~Kearney,
  and C.~Song, ``Bringing scientific workflow to the masses via {Pegasus} and
  {HUBzero},'' \emph{parameters}, vol.~13, p.~14.

\bibitem{runmycode}
V.~Stodden, C.~Hurlin, and C.~Perignon, ``{RunMyCode.org}: A novel
  dissemination and collaboration platform for executing published
  computational results,'' in \emph{E-Science (e-Science), 2012 IEEE 8th
  International Conference on}, 2012, pp. 1--8.

\end{thebibliography}
\end{document}